\documentclass[prb,twocolumn,showpacs,citeautoscript]{revtex4}

\usepackage{amsmath}
\usepackage{graphicx, epsfig}
\usepackage{color}
\usepackage{bm}
\usepackage{ulem}

\begin{document}

\bgroup\catcode`\#=12\gdef\hash{#}\egroup \newcommand {\pzoa} {$\mathrm{PbZrO_3}$}
\bgroup\catcode`\#=12\gdef\hash{#}\egroup \newcommand {\ptoa} {$\mathrm{PbTiO_3}$}
\bgroup\catcode`\#=12\gdef\hash{#}\egroup \newcommand {\pzta} {$\mathrm{PbTiO_3/PbZrO_3}$}
\bgroup\catcode`\#=12\gdef\hash{#}\egroup \newcommand {\pzo} {PZO}
\bgroup\catcode`\#=12\gdef\hash{#}\egroup \newcommand {\pto} {PTO}
\bgroup\catcode`\#=12\gdef\hash{#}\egroup \newcommand {\pzt} {PTO/PZO}
\bgroup\catcode`\#=12\gdef\hash{#}\egroup \newcommand {\sto} {$\mathrm{SrTiO_3}$}
\bgroup\catcode`\#=12\gdef\hash{#}\egroup \newcommand {\nso} {$\mathrm{NdScO_3}$}
\bgroup\catcode`\#=12\gdef\hash{#}\egroup \newcommand {\szo} {$\mathrm{SrZrO_3}$}
\bgroup\catcode`\#=12\gdef\hash{#}\egroup \newcommand {\kto} {$\mathrm{KTaO_3}$}
\bgroup\catcode`\#=12\gdef\hash{#}\egroup \newcommand {\kno} {$\mathrm{KNbO_3}$}

\bgroup\catcode`\#=12\gdef\hash{#}\egroup 
\newcommand{\ptl}{$P4mm$}      
\newcommand{\pth}{$Ima2$}      
\newcommand{\pz}{$A1a1$}       
\newcommand{\pztl}{$P4bm$}     
\newcommand{\pztm}{$P1a1$}     
\newcommand{\pztw}{$P2an$}     
\newcommand{\pzth}{$P2_1am$}   
\newcommand{\pzts}{$P2mm$}     


\newcommand{\pzc}{$C1c1$}      
\newcommand{\pztmc}{$P1c1$}    
\newcommand{\pztwc}{$Pnc2$}    
\newcommand{\pzthc}{$Pmc2_1$}  
\newcommand{\pztsc}{$Pmm2$}    


\newcommand{\Pz}{$\mathrm{P_{001}}$}
\newcommand{\Pxy}{$\mathrm{P_{110}}$}
\newcommand{\Rz}{$\mathrm{R_{001}}$}
\newcommand{\Rxy}{$\mathrm{R_{110}}$}
\newcommand{\Ay}{$\mathrm{A_{\bar{1}10}}$}

\title{Interplay of epitaxial strain and rotations in
PbTiO$_3$/PbZrO$_3$ superlattices\\ from first principles}

\author{Jeroen L. Blok}
\affiliation{Faculty of Science and Technology and MESA+ Institute for
Nanotechnology, University of Twente, 7500 AE Enschede, The Netherlands}
\author{Karin M. Rabe}
\affiliation{Department of Physics and Astronomy, Rutgers
University, Piscataway, New Jersey 08854-8019, USA}
\author{David Vanderbilt}
\affiliation{Department of Physics and Astronomy, Rutgers
University, Piscataway, New Jersey 08854-8019, USA}
\author{Dave H.A. Blank}
\affiliation{Faculty of Science and Technology and MESA+ Institute
for Nanotechnology, University of Twente, 7500 AE Enschede,
The Netherlands}
\author{Guus Rijnders}
\email{a.j.h.m.rijnders@utwente.nl}
\affiliation{Faculty of Science and Technology and MESA+ Institute for
Nanotechnology, University of Twente, 7500 AE Enschede, The Netherlands}

\date{\today}

\begin{abstract}
We present first-principles calculations of the structural phase
behavior of the [1:1] \pzta\ superlattice
and the \ptoa\ and \pzoa\ parent compounds
as a function of in-plane
epitaxial strain. A symmetry analysis is used to identify the
phases and clarify how they arise from an interplay between
different kinds of structural distortions, including
out-of-plane and in-plane polar
modes, rotation of oxygen octahedra around out-of-plane
or in-plane axes, and an anti-polar mode. Symmetry-allowed
intermode couplings are identified and used to elucidate the nature
of the observed phase transitions. For the minimum-period
[1:1] \pzta\ superlattice, we identify a sequence of three
transitions that occur as the
in-plane lattice constant is increased. All four
of the phases involve substantial oxygen octahedral rotations, and
an antipolar distortion is important in the high-tensile-strain phase.
Inclusion of these distortions is found to be crucial
for an accurate determination of the phase boundaries.
\end{abstract}


\maketitle

\section{Introduction}

Ferroelectric materials have long been the subject of intense
study both because of their fundamental scientific interest
and because their properties are promising for device applications.
Experimental and theoretical studies of the relationships between
their symmetries, structures and properties are essential for
the development of improved materials.  One of the most widely used
and thoroughly studied materials is $\mathrm{PbZr_xTi_{1-x}O_3}$ (PZT),
which in bulk shows an enhanced piezoelectric effect 
near the rhombohedral-tetragonal 
morphotropic phase boundary (MPB) around $x=0.5$.  Noheda et
al.\cite{Noheda2000} explained this anomalous response in
connection with their discovery of a narrow wedge of monoclinic
phase, bridging between tetragonal and rhombohedral phases at
the MPB.

Modern growth techniques such as pulsed laser deposition
and molecular-beam epitaxy now allow for the out-of-equilibrium
synthesis of atomic-scale
perovskite superlattices of arbitrary layer sequences, and 
control of the in-plane lattice constant via coherent epitaxy on
chosen perovskite substrates\cite{Rijnders2005}.
As the behavior of a bulk solid solution and a superlattice
having the same overall composition can in principle be quite
different, these developments offer an exciting opportunity for
the development of improved new materials. However, given the
enormous number of possible structures that can be constructed
in this way, it is
clear that theory has an important role to play in guiding the
search.

Not surprisingly, then, there has been a simultaneous explosive
growth in the application of first-principles theoretical
methods to the study of perovskite superlattice structures
\cite{junquera08}, specifically to the prediction of structure,
polarization and polarization-related properties such as the
dielectric and piezoelectric response.
Initially, the focus was on the electrostatic interaction
of the different layers, specifically on how a ferroelectric
instability in one constituent could induce polarization in the
other constituent \cite{neaton2003,johnston2005}, maintaining a uniform
displacement field throughout.\cite{wu2008} The
systems considered included superlattices of PbTiO$_3$ (\pto) and
PbZrO$_3$ (\pzo), specifically the [1:1] \pzt\ superlattice
with varying epitaxial strain\cite{Bungaro2004}.
However, it has long been recognized \cite{samara1975,Zhong1995-2}
that oxygen octahedron rotations play an important role in
determining structure, polarization and related properties in
ferroelectric perovskites.
In first-principles work, such antiferrodistortive instabilities
were identified in a number of ferroelectric
perovskites \cite{Ghosez1999}, and it was shown that octahedral
rotations in PZT play an important role in the morphotropic phase
boundary \cite{Fornari2001,kornev2006} and possibly in high-pressure
phases of \pto.\cite{kornev2005}  Most recently, combinations
of rotations at the interfaces were found in first-principles
investigations to generate improper ferroelectricity in \pto/SrTiO$_3$
superlattices\cite{bousquet2008} and in layered perovskite
compounds\cite{benedek2011,rondinelli2011b}. An overview of the
effect of rotational distortions on perovskites can be found in
Ref.~~\cite{rondinelli2011a}.

\begin{figure*}
\centering\includegraphics[width=6.0in]{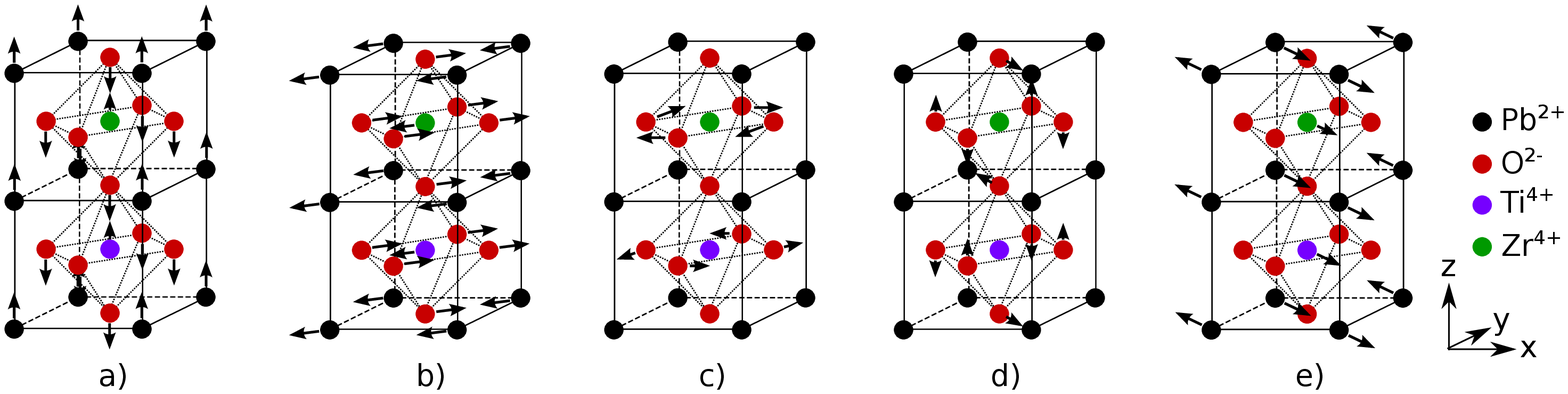}
\caption{Sketch of distortions of type (a) \Pz, (b) \Pxy, (c) \Rz,
(d) \Rxy\ and (e) \Ay\ in the \pzt\ superlattice.}
\label{distortions}
\end{figure*}

In the present work, we apply first-principles DFT methods to
determine the ground-state phases of the [1:1] \pzt\ superlattice
as a function of epitaxial strain, as well as those of pure \pto\
and \pzo\ for reference.  We go beyond previous work\cite{Bungaro2004} by
including the octahedral rotations and tilts (both of which
shall henceforth be referred to as ``rotations'') which
can be accommodated in a $\sqrt2\times \sqrt2 \times 2$
supercell; the types of distortions identified as playing an
important role
are shown in Fig.~\ref{distortions}. We find that octahedral rotations
are ubiquitous throughout the phase diagram of the \pzt\
superlattice, and have an significant influence on the polarization
and the locations of the phase boundaries.  In particular, they
help stabilize a monoclinic phase over a very broad range of
epitaxial strain.  We relate these results to the behavior of
the \pto\ and \pzo\ constituents and analyze the results
in terms of the symmetries of the various phases; our approach
is analogous to that for A-site superlattices reported in
Ref.~~\onlinecite{rondinelli2011b}.
Our results clarify the role played by the
coupling between these degrees of freedom in determining the
structure and polarization of the superlattice structure under
consideration. 

This paper is organized as follows. In Sec.~\ref{methods} we describe
the methods used to apply DFT to bulk \pto\ and \pzo\ and to the
\pzt\ superlattice. The results of these calculations are presented
is Sec.~\ref{results}, and a discussion is given in Sec.~\ref{discussion}.
We summarize and conclude in Sec.~\ref{conclusions}.

\section{methods}\label{methods}

\subsection{Computational details}
\label{sec:details}

The effect of rotations on the \pzt\ superlattice was studied using
density-functional theory (DFT) calculations within the local
density-approximation (LDA)\cite{Kohn1965,Ceperley1980}. The
calculations were performed using the Vienna Ab initio Simulation
Package (VASP)\cite{VASP4,VASP2}. The Brillouin zone was sampled
with a 4x4x4 Monkhorst-Pack grid\cite{Monkhorst1976}. The projector
augmented wave method (PAW)\cite{PAW1,PAW2} was used with an
800 eV plane-wave cutoff. The polarization for all systems was
calculated using the Berry-phase method\cite{King-Smith1993}
as implemented in VASP.

For comparison with the \pzt\ superlattice, we 
performed analogous studies of \pto\ and \pzo\ with the same epitaxial
constraints.  For all three
systems, the structure was first initialized by imposing selected polar
and/or rotational distortions,
not limited to those shown in Fig.~\ref{distortions},
and the system was then relaxed
subject to the constraint of fixed in-plane
lattice parameters, corresponding to the epitaxial boundary condition.
While the symmetry of some of the phases allowed for a tilting of the $c$
axis away from the normal to the $a$-$b$ plane, tests described
below in Sec.~\ref{sec:axistilt} show that this effect is small,
and the calculations reported here were all carried out with the $c$
axis kept parallel to $\hat{z}$.  The in-plane lattice parameters
were varied over the range
3.80\,\AA$\,<a<\,$4.10\,\AA\ for \pto,
3.90\,\AA$\,<a<\,$4.20\,\AA\ for \pzo\ and
3.90\,\AA$\,<a<\,$4.25\,\AA\ for the \pzt\ superlattices.

In order to consider the effects of rotations of
oxygen octahedra, the calculations were carried out on a 20-atom
$\sqrt2\times \sqrt2 \times 2$ unit cell.  This supercell allows
distortions with wavevectors at the $\Gamma$ point ($0,0,0$),
X point ($0,0,\pi/a$), R point ($\pi/a,\pi/a,\pi/a$) and $M$
point ($\pi/a,\pi/a,0$). In particular, it allows R point and M
point oxygen octahedron rotational distortions which are represented
most generally by $a^-b^-c^-$ and $a^-b^-c^+$ in the Glazer
notation.\cite{glazer1972}

In all cases considered,
we found that structures with P$_{100}>0$ had higher energies
than those with \Pxy$>0$.
Therefore in the following we limit our
consideration to structures with nonzero \Pz\ and/or \Pxy\ 
as shown in Figs.~\ref{distortions}(a-b).  Furthermore,
when comparing \pto\ and \pzo\ phases incorporating $R$-point
(out-of-phase) rotations to those with $M$-point (in-phase)
rotations, we always found the $R$-point patterns to be lower in
energy.  Similarly, with one exception that will be noted in due
course, we always found the rotations of the ZrO$_6$ and TiO$_6$
octahedra to have opposite signs in the \pzt\ superlattices.
In what follows, therefore, out-of-phase rotations of the two
layers stacked along $\hat{z}$ are to be understood unless
otherwise specified.
Rotations about [001] and [110] axes will be denoted by
\Rz\ and \Rxy\ respectively, as in Figs.~\ref{distortions}(c-d).
The 20-atom supercell also accommodates antipolar distortions of
the type shown in Fig.~\ref{distortions}(e), denoted as \Ay.
Structural energies for the \pto, \pzo\ and \pzt\ systems are
given relative to the energy of the corresponding ideal
high-symmetry $P4/mmm$ structure (i.e., with no rotations or
polarization) at a reference lattice constant of 3.99\,\AA.

\subsection{Small distortions}
\label{sec:small-distortions}

Typically we started our structural minimizations from an initial
geometry of relatively low symmetry, corresponding to a mixture of
several kinds of distortion, so as to be conservative in our
assumptions about what kind of structural distortions would
actually occur.  For example, in the \pto\ calculations
to be described below, we started from a system with the
\Pz, \Pxy, \Rz\ and \Rxy\ distortions
of Figs.~\ref{distortions}(a-d) present. After structural
relaxation, some of these components (especially the rotational
ones) would become very small but not disappear
completely. To determine whether these small distortions were
physically meaningful, or just a result of incomplete convergence
or related numerical difficulties, the distortions were manually
turned off (i.e., the structure was symmetrized) and then
the total energy was calculated as the distortions were added back
in in small increments.  If the resulting energy curve showed a
minimum away from zero distortion amplitude,
the distortion was taken to be physically meaningful.
This analysis proved useful, in particular, in clarifying
the phase diagram for \pto\ in Sec.~\ref{parent}.

\subsection{Location of phase boundaries}
\label{sec:boundaries}

As we shall show later, the case of the \pzt\ superlattice
presents a related difficulty.  Here, we find two second-order
transitions at which one or more order parameters go to zero
in an apparent square-root singularity.  In order to get a
more precise location for the critical epitaxial lattice constant
at which this occurs, we have carried out the following procedure.
We describe this method for the case of the transition from
the lower-symmetry \pztm\ space group to the higher-symmetry
\pztw\ space group with increasing epitaxial lattice constant
in the tensile-strain region of \pzt, but similar considerations apply
to the transition from \pztm\ to \ptl\ in the compressive-strain
region.

First, we chose a lattice constant near the transition, but far
enough away that the distortion leading to the low-symmetry phase was well
developed.  In our example, we chose $a$=4.125\,\AA, in the
\pztm\ phase but near the transition.  We symmetrized this structure
using the symmetry operations of the higher-symmetry phase,
constructed the eigenvector in displacement space pointing from the
symmetrized to the low-symmetry structure, and used this to define
a symmetry-breaking mode amplitude.  Next, for each lattice constant
near the transition (here for $a$=4.100 to 4.250\,\AA\ in increments
of 0.025\,\AA), we started from the relaxed high-symmetry structure
and added distortions corresponding to this eigenvector with several
amplitudes (and no further relaxation), fitting to an expression of the
form
\begin{equation}
\Delta E(u)=\frac{1}{2} \kappa u^2 + \frac{1}{4} \alpha u^4
\label{eq:simple}
\end{equation}
to the energy vs.\ mode amplitude $u$ (relative to $u$=0) as
computed from the first-principles calculations.
Not surprisingly, we found that $\kappa$ is strongly dependent
on in-plane lattice constant $a$, while $\alpha$ is not.
We then plotted $\kappa$ vs.\ $a$ (typically obtaining a nearly
linear behavior) and identified the critical lattice constant
$a_{\rm crit}$ as the one at which this curve crosses through zero.
As we shall see later, we found the transition to occur at
$a_{\rm crit}$=4.171\,\AA\ for the transition in this example.
As we shall also see, using the same analysis on the compressive
train side we found the transition from \pztm\ to \pztl\ to occur
at $a_{\rm crit}$=3.890\,\AA.

\subsection{Treatment of $c$-axis tilting}
\label{sec:axistilt}

As mentioned above, some of the space groups considered in this
work, specifically the monoclinic ones, allow by symmetry for
the $c$ axis to tilt away from the [001] direction, typically
by developing a small [110] component.  To study the effect of
such a shear strain, especially on the location of second-order
phase boundaries as described in the previous subsection,
we again made use of the distortion pattern leading from
the higher-symmetry tetragonal or orthorhombic state to the
lower-symmetry monoclinic one as obtained in Sec.~\ref{sec:boundaries}.
Again denoting the amplitude of this
normalized distortion by $u$, and starting from the higher-symmetry
structure at the lattice constant of interest, we carried out
calculations of changes in total energy and stress as a function
of $u$ and shear strain (axis tilt) $\eta$.  We then
parametrized the results to the first-principles calculations
using a generalized version of Eq.~(\ref{eq:simple}) taking the form
\begin{equation}
\Delta E(u,\eta)=\frac{1}{2} \kappa u^2 + \frac{1}{4} \alpha u^4 + \beta u
\eta + \frac{1}{2} \gamma\eta^2.
\end{equation}
Again $\kappa$ was the only parameter that was found to depend strongly
on $a$.  Setting $\partial\tilde{E}(u,\eta)/\partial\eta=0$ and
$\partial\tilde{E}(u,\eta)/\partial u=0$ within the model, we can
estimate both the shift of the phase boundary and the amount of
$c$-axis tilt.
For the phase boundary when the \pzt\ superlattice system goes
from the \pztm\ space group to the \pztw\ space group, for example,
this resulted in a shift of the critical in-plane lattice constant
by only 0.0016\,\AA, to 4.173\,\AA,
leading to small shear strains near the phase boundary ($\sim$0.2\%).
Similar results were found for other phase boundaries between
monoclinic and tetragonal or orthorhombic phases.

The impact
of the tilting of the $c$ axis appears to be less profound in
our study than in that of Ref.~~\onlinecite{Bungaro2004}, in
which octahedral rotations were not included.  Indeed, we find
it to be small enough that it will not change any of the results
significantly, and so we have not included it in the results
to be presented in the next section.

\section{Results}\label{results}

\subsection{Parent phases}
\label{parent}

\begin{figure}
\centering\includegraphics[width=3.0in]{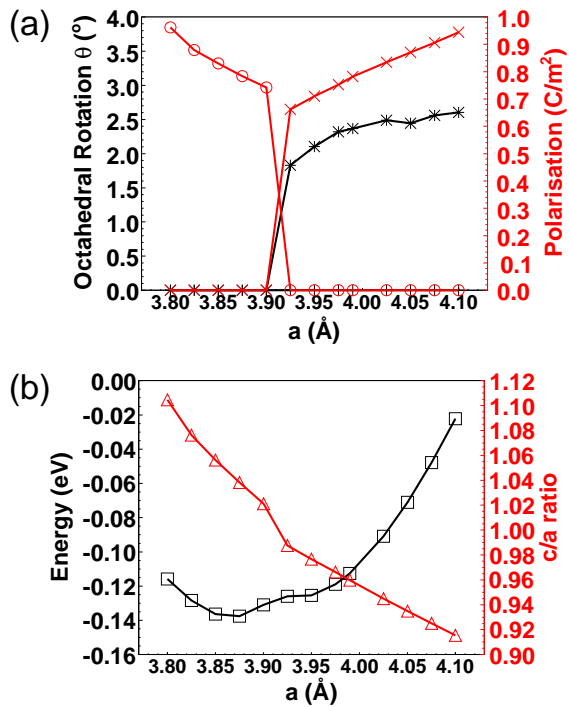}
\caption{(Color online) Calculated properties of \pto\ as a
function of epitaxially constrained in-plane lattice constant.
(a) Minimum-energy structure:
octahedral rotations (black) of type \Rxy\ ($\ast$);
polarizations (red) of type \Pz\ ($\circ$) and \Pxy\ ($\times$).
(b) Total energy (black squares) and $c/a$ ratio (red triangles).}
\label{pto}
\end{figure}

The results of the DFT calculations on \pto\ are given in Fig.~\ref{pto}.
For in-plane lattice parameters less than about
$a=3.90$\,\AA, we find the stable structure to be the tetragonal
\ptl\ structure (space group 99) in which only \Pz\ is present
(ferroelectric ``$c$ phase,''
5-atom primitive cell), as in Fig.~\ref{distortions}(a).
For larger lattice parameters, we see a
transition at $a=3.91$\,\AA\ to an orthorhombic \pth\ structure
(space group 46, 10-atom primitive cell) with nonzero \Pxy\ 
accompanied by an \Rxy\ rotation of $\sim$2$^\circ$,
combining the distortions of Figs.~\ref{distortions}(b) and (d).
Since there is
no group-subgroup relationship between these two phases, the
transition must be first order.
These results are in rather good
agreement with the previous first-principles calculations of
Refs.~~\onlinecite{Bungaro2004} and ~\onlinecite{Dieguez2005}
(assuming mixed domain phases are not considered), even
though those works did not include the octahedral rotations.
This suggests that the small rotations that we predict on the
higher-$a$ side of the transition do not have a very profound effect
on the transition in \pto.
We also note that all of the first-principles results disagree
with the zero-temperature prediction of a phenomenological
analysis\cite{Pertsev1998} where a window of $r$ phase
(\Pz\ and \Pxy\ both present) was predicted at $T$=0.

\begin{figure}
\centering\includegraphics[width=3.0in]{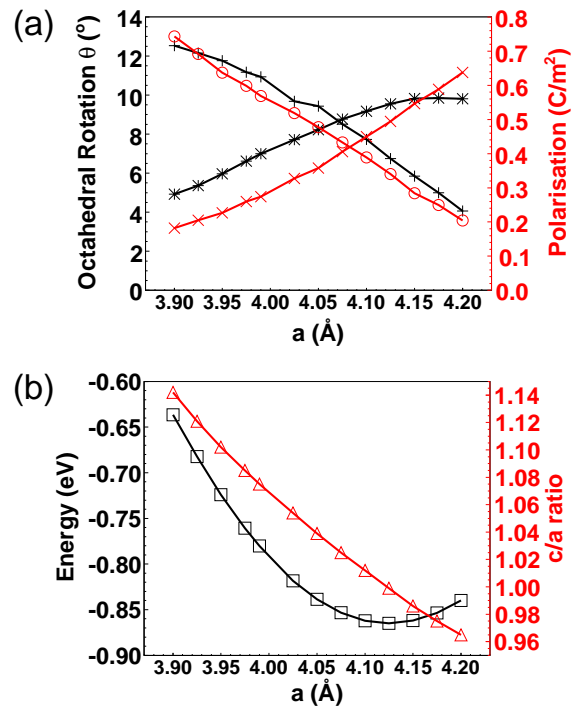}
\caption{(Color online) Calculated properties of \pzo\ as a
function of epitaxially constrained in-plane lattice constant.
(a) Minimum-energy structure:
octahedral rotations (black) of type \Rxy\ ($\ast$) and \Rz\ ($+$);
polarizations (red) of type \Pz\ ($\circ$) and \Pxy\ ($\times$).
(b) Total energy (black squares) and $c/a$ ratio (red triangles).}
\label{pzo}
\end{figure}

As is well known, bulk \pzo\ has a complicated
$Pbam$ 40-atom unit cell built from a
$2\sqrt{2}\times\sqrt{2}\times2$ enlargement of the parent
perovskite unit cell \cite{Singh1995}. The structure is described
by a combination of several distortions, the largest being an
R-point oxygen octahedron rotation (of type $a^-a^-c^0$ in Glazer
notation) and an antiferroelectric
($\pi/2a,\pi/2a,0$) $\Sigma$-line mode\cite{fujishita1984}. The
latter is not allowed within the supercell used in our
calculations. However, in bulk PbZrO$_3$, the pattern of atomic
displacements for this mode is dominated by antipolar displacements
of the Pb atoms in the [001] plane, and it is unlikely to survive
in our [1:1] superlattice, in which the Pb atoms are shared with
the \pzo\ layer.
Therefore, since our immediate goal is to
provide a reference for the \pzt\ superlattice calculations, we
continue to restrict our calculations to structures realizable
within our 20-atom simulation cell.  After considering a wide
variety of structures consisting of different combinations of
polar and rotational distortions, we find that the
monoclinic \pz\ phase (space group 9, \pzc\ in the
conventional setting, 10 atoms/cell), is lowest in
energy throughout the range of in-plane lattice constants
considered, as shown in Fig.~\ref{pzo}.  All four of the distortions
\Pz, \Pxy, \Rz, and \Rxy\ shown in Figs.~\ref{distortions}(a-d)
are present in this phase.

\subsection{\pzt\ superlattice}
\label{superlattice}

\begin{figure}
\centering\includegraphics[width=3.0in]{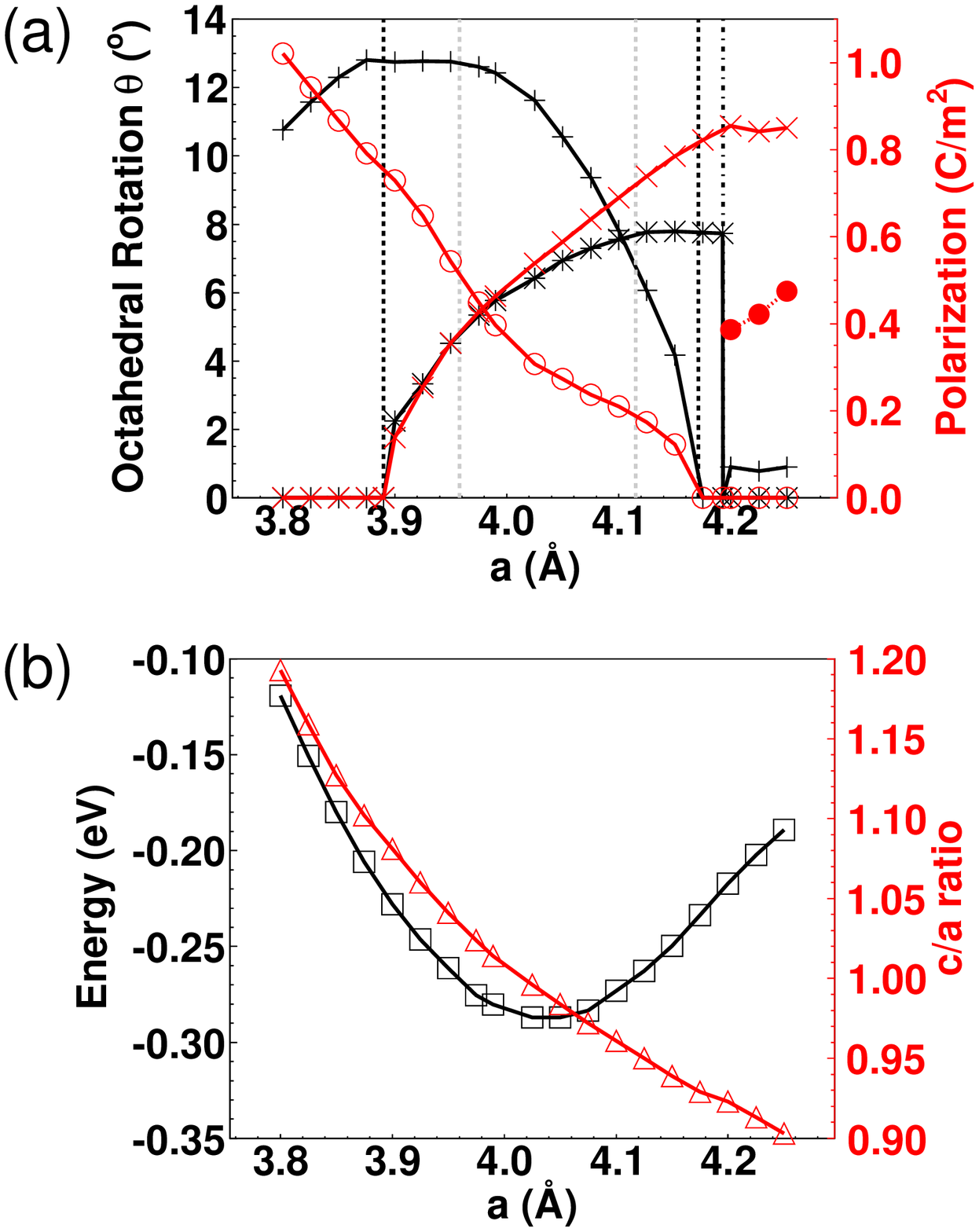}
\caption{(Color online) Calculated properties of \pzt\ superlattice as a
function of epitaxially constrained in-plane lattice constant.
(a) Minimum-energy structure:
octahedral rotations (black) of type \Rxy\ ($\ast$) and \Rz\ ($+$);
polarizations (red) of type \Pz\ ($\circ$) and \Pxy\ ($\times$);
antipolar distortions of type \Ay\ (filled red circles).
(b) Total energy (black squares) and $c/a$ ratio (red triangles).
Heavy and light vertical dotted lines indicate phase boundaries
determined from our work, and from Ref.~~\onlinecite{Bungaro2004},
respectively.}
\label{pzt}
\end{figure}

We now turn to the case of the \pzt\ superlattice.
Compared to the relatively simple behavior for pure \pto\ and pure \pzo,
the phase sequence for the \pzt\ superlattices presented in
Fig.~\ref{pzt} is quite rich.

We first remind the reader of the previous results of Bungaro and Rabe
\cite{Bungaro2004}, who confined their considerations to structures
that do not include octahedral rotations.  Within this constraint,
these authors found a sequence of two phase transitions, first from
a tetragonal $P4mm$ structure (``$c$ phase'') for
compressive strains $a< 3.96$\,\AA to a monoclinic
$Cm$ structure (``$r$ phase'') at intermediate
strains, and finally to an orthorhombic $Amm2$ structure
(``$aa$ phase'') for tensile strains $a> 4.12$\,\AA. 
Our results, presented in Fig.~\ref{pzt}, show a qualitatively
similar behavior for the polarization, although all of the
phases we observe include octahedral rotations, and the phase
boundaries (vertical dashed lines in Fig.~\ref{pzt}) have shifted
to make the region of monoclinic phase wider.  We give a detailed
presentation of our results in the following, first in the region
of compressive strain, and then in the tensile region, where the
behavior is especially subtle.

\subsubsection{Compressive strain region}
\label{compressive}

Starting at the smaller lattice constants $a<3.890$\,\AA, we
find a tetragonal \pztl\ structure (space group 100)
in which \Pz\ and \Rz\ are both present
(20-atom primitive cell).  The octahedral
rotation angles for \Rz\ in the \pto\ and \pzo\ planes are quite
different, with the
\pto\ rotations remaining rather small, $\sim$1$^\circ$.
(Throughout the figure, only the \pzo\ rotations are plotted.)
When the lattice constant exceeds 3.890\,\AA, we find that \Pxy\ and \Rxy\
turn on simultaneously in a second-order transition,
leading to a monoclinic \pztm\ structure (space group 7,
\pztmc\ in the conventional setting, 20-atom primitive cell).
This structure is very similar to the \pz\ monoclinic
structure obtained for \pzo\ in Fig.~\ref{pzo}, although the
inequivalence of the \pto\ and \pzo\ layers makes the
\pztm\ structure more complicated.
The quoted value of $a_{\rm crit}$=3.890\,\AA\ was obtained
using the approach described in Sec.~\ref{sec:boundaries},
but applied in the compressive-strain region, and using the
distorted structure at $a$=3.925\,\AA\ to define the
distortion eigenvector.

\subsubsection{Tensile strain region}
\label{tensile}

Next, as the lattice constant increases and approaches a value
of about 4.19\,\AA, we find that the amplitudes of \Pz\ and \Rz\
diminish and approach zero together.  Again using the method of
Sec.~\ref{sec:boundaries}, we find that these mode amplitudes
disappear in a second-order transition at $a$=4.171\,\AA,
leading to an orthorhombic \pztw\ phase (space group 30,
\pztwc\ in the conventional setting) in which only \Pxy\ and \Rxy\
are present (20-atom primitive cell).  This structure
is similar to the \pth\ phase seen at larger lattice
constants for \pto\ in Fig.~\ref{pto}, except that the \pto\ and \pzo\ planes
are inequivalent.

However, at a marginally larger lattice constant ($a$=4.193\,\AA),
a first-order transition occurs to an orthorhombic \pzth\ phase
(space group 26, \pzthc\ in the conventional setting, 20-atom
primitive cell).  The value of $a_{\rm crit}$=4.193\,\AA\ was
obtained from the crossing of curves of computed energy vs.\
lattice constant as computed in the \pztm\ and \pzth\
symmetries.  The \pzth\ phase is one in which \Rxy\ disappears,
\Pxy\ is approximately unchanged, and \Ay\ and \Rz\ distortions
appear.  \Ay, denoted
by the solid red dots in Fig.~\ref{pzt}, consists of displacements
(largest for Pb, then Ti, then Zr, but also involving O atoms)
that are parallel and antiparallel to ($\bar{1}$10), patterned
in such a way that no net polarization develops along this axis,
as shown in Fig.~\ref{distortions}(e).  
(Such antipolar distortions are also symmetry-allowed in the
monoclinic \pztm\ phase, but we find them to be very small, of
order 0.02\,\AA, suggesting that they are induced there by
intermode couplings as will be discussed in Sec.~\ref{sec:disc-sup}.)
Contrary to the usual case, the signs of \Rz\ are the same in the \pto\
and \pzo\ planes, with the rotations in both layers being small
($\sim$1$^\circ$).  The fact that the displacements associated
with \Ay\ are much larger than those associated with \Rz\ suggests
that \Ay\ and \Pxy\ constitute the primary instabilities;
when these modes freeze in, they induce an \Rz\ distortion,
as will become clear from the symmetry
analysis to be given in Sec.~\ref{discussion}.

The fact that the second-order and first-order
transitions occur at almost the same lattice constant has to be
regarded as a coincidence.  The window of \pztw\ phase found in
our calculation, 4.171\,\AA$<a<$4.193\,\AA, is in fact so narrow
that we cannot be confident of its existence at all.
Small changes in the technical details of the
calculations, such as the exchange-correlation potential or choice
of pseudopotentials, could erase or widen this window of stability.
The inclusion of $c$-axis tilting, as discussed in
Sec.~\ref{sec:axistilt}, also tends to reduce or eliminate the
window of \pztw\ phase.
We can only conclude that the \pztw\ phase is competitive
with the other two phases near the transition from the
\pztl\ to the \pzth\ phase, and may appear in a narrow window
between them.  
If the \pztw\ phase is not in fact present, then there would
be a direct transition from \pztl\ to \pzth. (Despite the presence
of a group-subgroup relation, the transition would be first order
because of the discontinuity in \Rxy.)

\section{Discussion}\label{discussion}

This system illustrates several interesting aspects of complex
perovskite oxides.  Perovskite oxides generally are characterized
by multiple unstable modes in the cubic structure, which couple
and compete to produce low-energy states. The instabilities
also each separately couple to strain. Changing strain not only
produces or eliminates instabilities, but it can also change
the balance of competition and produce first-order phase transitions
between distinct phases without a group-subgroup relation.  Here,
we see that combining two different perovskite constituents,
each with characteristic unstable modes and strain dependence,
can lead to a great increase in richness of the observed phases.

\subsection{Parent phases}

In the present case, the computed phases in the two constituent
materials are relatively simple.  The dominant instability in
cubic \pto\ is the zone-center polar mode, which couples to
strain so that elongation is favored along the polar direction.
Compressive epitaxial strain thus favors \Pz , while tensile epitaxial
strain favors \Pxy .  In contrast to the proposed pressure-induced
tetragonal-monoclinic-rhombohedral phase sequence,\cite{wu2005}
the absence of a monoclinic phase
with rotating polarization, or equivalently the first-order
character of the transition, can be understood as arising from
the strong coupling of the polarization to tetragonal strain,
as demonstrated by the quartic anisotropy of the polarization	 
energy given in Table III of Ref.~~\onlinecite{Dieguez2005}. 
Additional symmetry breaking for tensile strain results from the 
rotational instability \Rxy . While compressive strain is expected to 
favor the rotational distortion \Rz , no instability occurs in the range 
of strain considered here.

In contrast, cubic \pzo\ has multiple instabilities that occur
together, yielding a monoclinic phase under the constraint
of the supercell used for the first principles calculations.
This phase is strongly favored and appears for all values of
strain considered, with compressive strain favoring the
modes \Pz\ and \Rz\ and tensile strain favoring the modes
\Pxy\ and \Rxy, but not with such a strong dependence as to
eliminate any of the modes at the highest strains considered. 
Also in contrast to \pto, the
polarization freely rotates in the ($1\bar10$) plane with changing
epitaxial strain; this was already apparent in the first-principles
results for strained \pzo\ neglecting rotations, which included
an intermediate monoclinic phase between the phases with normal
and in-plane polarization\cite{Bungaro2004,Dieguez2005}.

\subsection{\pzt\ superlattice}
\label{sec:disc-sup}

We now turn to the superlattice phase sequence. The point of
closest resemblance to the phases of the constituents is the
monoclinic phase for intermediate strain, which shows the rotating
polarization and accompanying rotations observed in \pzo.  However,
the strain dependence of the mode amplitudes is much stronger.
With increasing compressive strain,  \Pxy\ and \Rxy\ disappear, yielding 
a phase with only \Pz\
and \Rz . With increasing tensile strain \Pz\ and \Rz\
disappear, yielding a phase with only \Pxy\ and \Rxy, which at
slightly larger in-plane lattice constant undergoes a first-order
transition to a phase dominated by \Pxy, \Ay\ and \Rz.

A symmetry analysis of the Taylor expansion of the energy around a 
high-symmetry reference structure in the amplitudes of the relevant 
distortion modes is useful for understanding the mode contents of the 
various phases observed. For the pure compounds, the high-symmetry reference structure is taken as the tetragonal P4/mmm structure with 5 atoms per cell and all atoms at high-symmetry positions.
For the superlattice, the high-symmetry reference structure is the 
10-atom P4/mmm structure. We consider the distortion modes shown in 
Figure 1; for the
pure compounds there is an extra translation symmetry relating the two B 
cations and the modes shown are either even (\Pz, \Pxy, \Ay ) or odd 
(\Rz, \Rxy ) under this translation.

For pure \pto , the analysis is simple. For compressive strain, the \Pz\ 
mode is unstable and is the only mode included in the lowest-energy 
phase. For tensile strain, the lowest-energy phase includes both \Pxy\ 
and \Rxy\ distortions. In previous work \cite{Bungaro2004}
we have seen that the \Pxy\ mode is unstable for tensile strain.
\Rxy\ is thus nonzero either 
because it is independently unstable, as would be favored by tensile 
strain, or because the biquadratic coupling between \Pxy\ and \Rxy\ is 
cooperative and large enough to destabilize \Rxy .

For pure PZO, the four modes \Pz , \Pxy , \Rz and \Rxy are included
in the lowest-energy phase. It is
reasonable to assume that the polar distortions \Pz\ and \Pxy\ are 
unstable throughout this range, with instability in \Rz\ promoted by 
compressive strain and the instability in \Rxy\ promoted by tensile 
strain. However, the rotational modes need not be independently 
unstable.
Biquadratic coupling to the polar modes 
could produce the effective instability.
Furthermore, there is a fourth 
order invariant \Rxy\Pxy\Rz\Pz\ in the energy expansion that ensures 
that if three of the four distortions are present, the fourth will be 
induced through an improper mechanism even if it is not unstable.

The effect of this fourth-order coupling is particularly striking in the 
superlattice, as it is responsible for the coupled 
appearance/disappearance of \Pxy\ and \Rxy\ at $a$=3.890\,\AA\ and the 
coupled appearance/disappearance
of \Pz\ and \Rz\ at $a$=4.171\,\AA . The fact that these modes appear 
and disappear together might at first seem surprising, since in the
high-symmetry reference structure, these two distortions
have different symmetry and the introduction of one mode into
the reference structure would not induce the other.  However,
with the lowered symmetry due to the presence of \Pz\ and \Rz ,
the two modes \Pxy\ and \Rxy\ belong to the same irreducible 
representation and
therefore are expected to occur together.  This is formally equivalent 
to the presence of the fourth-order invariant \Rxy\Pxy\Rz\Pz ,
discussed above.
For example, in the compressive-strain phase with \Pz\ and \Rz , if
an instability with respect to \Pxy\ or \Rxy\ develops,
\Rxy\ or \Pxy\ will be
linearly induced through this fourth-order term, producing the
intermediate monoclinic phase
and ensuring that the two modes appear/disappear together.

It is in the extension of this symmetry analysis to include the \Ay\ 
distortion that differences between the pure compounds and the 
superlattice arise.
In the pure compounds, the \Ay\ distortion is a symmetry-breaking 
distortion that would appear only if it were unstable, which is not the 
case for the strain range considered. 
In the superlattice, 
there are two third-order couplings \Rz \Pxy \Ay\ and \Rxy \Pz 
\Ay, not allowed in the pure compounds due to the additional 
translation symmetry. In the intermediate-strain monoclinic phase, 
both of these terms would linearly induce an \Ay\ distortion, which is indeed
present at small amplitude in the computed structure for this phase.

In contrast to its minor role in the intermediate-strain monoclinic
phase, \Ay\ appears to be a strong independent instability in the 
superlattice
at higher tensile strain.  This can explain the first-order
transition at $a$=4.193\,\AA .
The approximate continuity of \Pxy\ through the
first-order transition suggests that \Pxy\ is by far the dominant
instability at higher tensile strain. Energy lowering from this
polar phase can be produced in two ways: through freezing in an
\Rxy\ instability to obtain the $Pnc2$ phase, or by freezing in an
\Ay\ instability, which induces a small \Rz\ distortion through
the third-order invariant \Ay\Rz\Pxy\ and yields the $Pmc2_1$ phase.
The key to the first-order transition is the competition of these
two instabilities in the presence of \Pxy.
 
\subsection{Growth of \pzt\ and related superlattices}
We have attempted the growth of [1:1] \pzt\ superlattices using
pulsed laser deposition (PLD), but without notable success.
Typically, reflection high-energy electron diffraction
(RHEED) indicates that an initial layer-by-layer growth rapidly
gives way to a 3D growth mode.  Our best result to date is
for growth on a \nso\ (NSO) substrate, where we observed
five successful coherent repetitions of the [1:1] \pzt\ unit
before the 3D growth mode took over. These results were obtained
using the deposition settings that were optimized for \pto\ by
Catalan et al.~\cite{catalan2006} The settings for \pto\ were
also used for the \pzo\ growth.

Previous work on \sto/\szo~\cite{christen1998} and
\kto\-/\kno~\cite{specht1998} systems has shown that the synthesis of B-site
superlattices is possible.
We speculate that the development of new substrates (or new
substrate treatments) might help extend the layer-by-layer
growth of \pzt, especially for larger lattice parameters
($a$=4.0-4.2\,\AA).  One option could be the new treatment
developed for scandates (including NSO) by Kleibeuker et
al.~\cite{kleibeuker2010}

Even if the extended growth of \pzt\ superlattices remains elusive,
this work may still provide useful guidelines for designing new
ferroelectric materials based on B-site superlattices. This can
be done either by creating
new low-symmetry phases by making use of fourth-order coupling
terms such as \Pz\Rz\Pxy\Rxy, or by inducing improper
ferroelectricity via a third order term such as \Rz \Pxy \Ay\ or
\Rxy \Pz \Ay.  For example, in the latter scenario, a system with an \Ay\
instability and an \Rz\ or \Rxy\ rotational instability should
be an improper ferroelectric.  For A-site
superlattices, recent work on the \pto/\sto\
system~\cite{bousquet2008} demonstrates that it is indeed possible
to synthesize superlattices that induce improper ferroelectricity,
even if the details of the symmetry analysis are somewhat different
in that case as is described by Rondinelli and Fennie\cite{rondinelli2011b}.

\section{Conclusions}\label{conclusions}

In summary, we have presented the results of first-principles
calculations of the structural phase transitions in
[1:1] \pzt\ superlattices, as well as in the parent compounds, as
a function of in-plane epitaxial strain. A symmetry analysis was
used to clarify the kinds of structural distortions occurring
in each phase and to illuminate the nature of the transitions
between phases.

While the \pzo\ system remains in a single monoclinic phase over
the whole region of lattice constants studied, and the \pto\
system has a single transition from a tetragonal to an orthorhombic
phase, we find a significantly more complex phase sequence for
the \pzt\ superlattice.  Specifically, with increasing lattice
constant, we find a sequence of three transitions, first from a
tetragonal to a monoclinic phase, then to a first orthorhombic
phase, and finally to a second orthorhombic phase.  The first
orthorhombic phase occurs over such a narrow range of in-plane
lattice constant that we cannot be confident of its existence
in real \pzt\ superlattices.  The first three of these phases
all contain strong polar and octahedral-rotation distortions,
while the fourth has strong in-plane polar order, substantial
antipolar distortions, and weak rotations.

Our results indicate that polarizations and rotational distortions
are strongly coupled in the \pzt\ superlattice.  For example,
for the monoclinic phase, in which four kinds of mode distortions
are present simultaneously, we find these modes to be coupled
such that the presence of any three of them will automatically
induce the fourth, giving rise to a surprisingly rich behavior.
Thus, we find that the inclusion of the octahedral rotations is
crucial for an accurate determination of the phases and phase
boundaries.

While the synthesis of short-period perovskite superlattices
has the potential to lead to a wealth of new functional materials
with improved properties, a thorough understanding of the
interplay between epitaxial strain and spontaneous structural
distortions will be needed to guide the search for such
materials.  The present work is a step along the way to this
goal, and we hope that it might help identify new materials
with enhanced ferroelectric and piezoelectric properties for
next-generation applications.

\acknowledgments

The work was supported by ONR grants N00014-05-1-0054 and N00014-09-1-0300.
This work is part of the research program of the Foundation for
Fundamental Research on Matter [FOM, financially supported by
the Netherlands Organization for Scientific Research (NWO)]. The
technical assistance of V.R. Cooper in the early states of the
work is gratefully acknowledged.

\bibliography{ferrobib}

\end{document}